%% file: main.tex
\documentclass[acmsmall]{acmart}

\acmConference[CSCW '23]{CSCW '23}{Minneapolis, MN}{USA}
\acmBooktitle{CSCW '23: The 26th ACM Conference on Computer-Supported Cooperative Work and Social Computing, Virtual}
\input{commands}

\newcommand{\sys}{Judgment Sieve}

\begin{document}

\title{\sys: Reducing Uncertainty in Group Judgments through Interventions Targeting Ambiguity versus Disagreement}


\author{Quan Ze Chen}
\email{cqz@cs.washington.edu}
\affiliation{%
  \institution{University of Washington}
  \city{Seattle, WA}
  \country{USA}}

\author{Amy X. Zhang}
\email{axz@cs.washington.edu}
\affiliation{%
  \institution{University of Washington}
  \city{Seattle, WA}
  \country{USA}}

\ccsdesc[500]{Human-centered computing~Collaborative interaction}

\keywords{crowdsourcing; annotation; ambiguity; calibration}

\input{s_abstract}

\maketitle

\input{s_intro}

\input{s_related}

\input{s_design}

\input{s_experiments}

\input{s_discussion}

\input{s_conclusion}


\bibliographystyle{ACM-Reference-Format}
\bibliography{biblio}


\end{document}

%% file: commands.tex
\usepackage{soul}
\usepackage{subcaption}
\usepackage{multirow}
\usepackage{csquotes}

\newcommand{\Comment}[1]{}

\usepackage{graphicx}
\usepackage{enumitem}
\setlist[description]{leftmargin=2\parindent,labelindent=2\parindent,rightmargin=2\parindent}

\setlist[itemize]{leftmargin=*}
\setlist[enumerate]{leftmargin=*}


%% file: s_abstract.tex
\begin{abstract}
When groups of people are tasked with making a judgment, the issue of uncertainty often arises. 
Existing methods to reduce uncertainty typically focus on iteratively improving specificity in the overall task instruction.
However, uncertainty can arise from multiple sources, such as ambiguity of the item being judged due to limited context, or disagreements among the participants due to different perspectives and an under-specified task.
A one-size-fits-all intervention may be ineffective if it is not targeted to the right source of uncertainty.
In this paper we introduce a new workflow, \sys, to reduce uncertainty in tasks involving group judgment in a targeted manner. By utilizing measurements that separate different sources of uncertainty during an initial round of judgment elicitation, we can then select a targeted intervention adding context or  deliberation to most effectively reduce uncertainty on each item being judged. We test our approach on two tasks: rating word pair similarity and toxicity of online comments, showing that targeted interventions reduced uncertainty for the most uncertain cases. In the top 10\% of cases, we saw an ambiguity reduction of 21.4\% and 25.7\%, and a disagreement reduction of 22.2\% and 11.2\% for the two tasks respectively. We also found through a simulation that our targeted approach reduced the average uncertainty scores for both sources of uncertainty as opposed to uniform approaches where reductions in average uncertainty from one source came with an increase for the other. 
\end{abstract}

%% file: s_intro.tex
\section{Introduction}

\begin{figure}[t]
    \centering
    \includegraphics[scale=0.29]{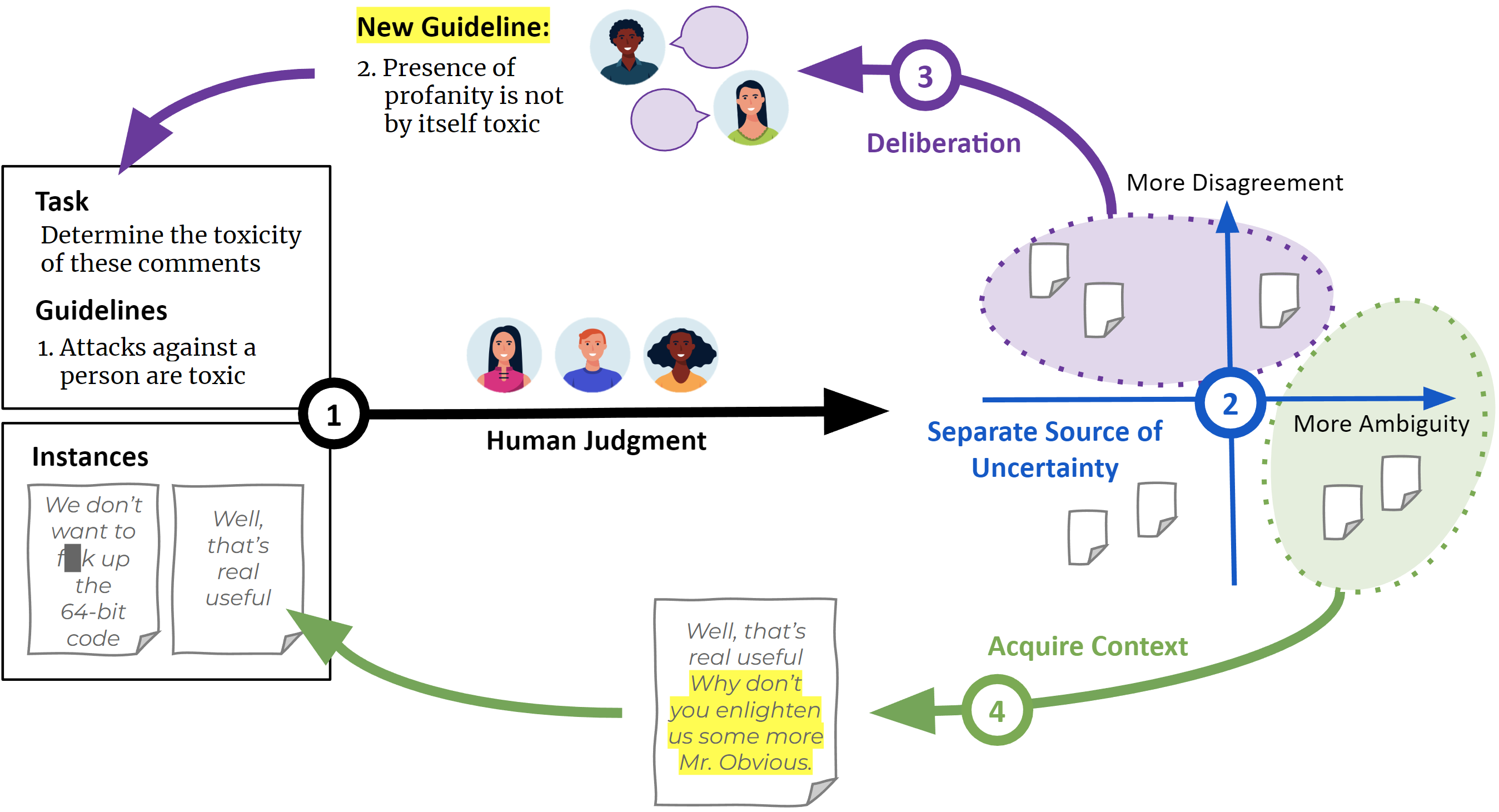}
    \caption{A high-level overview of the workflow: (1) Human judgments are collected using an annotation tool that quantifies distinct sources of uncertainty; (2) For each instance, scores that correspond to sources of uncertainty (i.e., \textit{ambiguity} and \textit{disagreement}) are computed; (3) Instances with more disagreement are given the \textsc{deliberation} intervention to resolve disagreements, producing new guidelines; (4) Context is collected for instances with more ambiguity and incorporated into the instance.}
\label{fig:workflow-overview}
\end{figure}

Uncertainty is an unavoidable challenge in many tasks that involve making judgments on items. In particular, judgments that involve groups of people must grapple with uncertainty often, as uncertainty in the group setting can arise from both uncertainty experienced by individuals in the group as well as uncertainty at the group level.
Individuals in the group may each feel some level of uncertainty due to the \textit{ambiguity} of the item they are judging, making it hard to personally decide on a judgment.
At the same time, even if individuals are certain in their personal judgments, group uncertainty can still arise due to \textit{disagreement} between members of the group, which can come from differences in perspectives of the group members that have not been addressed via a specification in the task instructions.

For example, a group of community moderators determining whether a post should be taken down for being harmful may face uncertainty due to ambiguity in the language used and the poster's intent~\cite{pavlopoulos-etal-2020-toxicity}. At the same time, differences in background and culture~\cite{Jiang2021UnderstandingIP} mean that members of the community may often disagree on what even is harmful~\cite{Baker2020TheCO} and what actions should be taken as a consequence~\cite{Atreja2022WhatIT}. Similarly, in the education setting, teaching assistants and instructors are often faced with uncertainty when grading open ended assignments. While the goal of grading is to evaluate a student's level of understanding, a poorly designed assignment question may lead to an answer that does not clearly demonstrate the student's understanding one way or the other. Separately, if a shared grading rubric doesn't specify what to do in a particular case, different graders may end up relying on personal judgment, creating disagreement and inconsistencies~\cite{singh-2017-gradescope}. 

Failure to account for uncertainty during the process of human annotation can lead to unreliable and inconsistent measurements~\cite{welty2019-metrology} even in domains involving expert judgments~\cite{cortes2021-inconsistency-conference}. Additionally, biases resulting from different backgrounds and perspectives of individuals in the group can also create biased group judgments if not properly accounted for~\cite{sap2019-risk,sap2021-annotators}. Due to these observations, in many areas, approaches and processes have been developed to measure uncertainty in data in order to discard unreliable judgments ~\cite{bhatt-etal-2021-transparency,gordon2021-deconvolution,prabhakaran-etal-2021-releasing} or to create systems that can make use of disaggregated data~\cite{fornaciari-etal-2021-beyond,leonardelli-etal-2021-agreeing,wang-2021-label-disagreement}. While accounting for existing uncertainty is important for building more robust processes, when it comes to actually making decisions, in many cases intervention to reduce uncertainty becomes necessary.

One intervention to reduce uncertainty tackles the ambiguity in the item being judged by providing additional context to help make a decision.
For example, providing information such as the parent post of a comment has been shown to sometimes result in an opposite judgment of the toxicity~\cite{pavlopoulos-etal-2020-toxicity}. 
However, context is complex, and collecting the full scope of relevant context for all cases can be difficult~\cite{solorio-etal-2014-sockpuppet}. Thus, it is often infeasible to collect context for all instances ahead of time. 
Another way to reduce uncertainty tackles the disagreement between individual judgments in the group. 
Several methods have been proposed to address disagreement by adding greater specification to the task, such as through the use of anchoring examples~\cite{chen2021-goldilocks} to ground task understanding and using measured uncertainty to find unclear guidelines~\cite{Manam2019-TaskMateAM}. Other methods tackle the underlying issue of differing perspectives, where deliberation has been shown to improve consensus~\cite{Schaekermann2018-resolvable,Chen2019-CiceroMC,Fan2020-DigitalJA}. However, these interventions also come at a high cost, often requiring a synchronous collaboration process.  

Not only are these interventions costly but applying an intervention meant to address one form of uncertainty when the cause of uncertainty lies elsewhere may lead to wasted effort. For instance, prior work has found that resolution of disagreements via deliberation can fail when the context is ambiguous or missing~\cite{Schaekermann2018-resolvable}.
On the other hand, more context may not be helpful if group members are certain about their judgment but still need to resolve disagreements. Instead of applying all of these costly interventions in every case that presents uncertainty,  if we can measure and distinguish the sources contributing to group uncertainty for each case, then it would be possible to select a more targeted and effective intervention on a per-case level. 

In this paper, we present a new workflow, \sys, for efficiently reducing uncertainty in group judgments.
\sys involves a decision process that selects a targeted intervention based on the types of uncertainty observed during the initial annotation of each item (Figure~\ref{fig:workflow-overview}). When individual uncertainty is detected, we focus on acquiring more context to reduce ambiguity in the item; when disagreement between annotators is detected, we focus on engaging annotators in deliberation to reconcile their diverging perspectives and better specify the task instructions.

We make the following contributions in this paper:
\begin{itemize}
    \item We present \sys, a workflow for reducing uncertainty in group judgment scenarios by utilizing measurements related to ambiguity and disagreement for each instance. We also provide a prototype implementation of this workflow for scalar rating tasks.
    \item We conduct annotations on two scalar rating task domains: word pair similarity rating (\textit{wordsim}) and toxicity rating (\textit{toxicity}), and verify that adding context and deliberation are \textbf{effective interventions for reducing ambiguity and disagreement respectively}. 
    \begin{itemize} 
        \item In the top 10\% most ambiguous cases, we observed a 21.4\% (\textit{wordsim}) and 25.7\% (\textit{toxicity}) reduction in ambiguity by introducing context. 
        \item Similarly for the top 10\% highest disagreement cases, we saw a 22.2\% (\textit{wordsim}) and 11.2\% (\textit{toxicity}) reduction in disagreement by introducing guidelines created from deliberation. 
    \end{itemize}
    \item However, we also observed that a broad application of interventions over all items can increase uncertainty in some circumstance, where adding context increased disagreement by 2.06\% (\textit{wordsim}) and 3.54\% (\textit{toxicity}). 
    \item We conduct a simulation experiment to evaluate the targeted intervention aspect of \sys~which selects an intervention based on the type of uncertainty measured in the initial annotation. We find that targeted selection of interventions applied to the most uncertain examples resulted in reductions in the overall means of both uncertainty sources as compared to a uniform approach where reductions in one source of uncertainty came with an increase in the other. Though, we do note that when including instances where our dynamic approach did not assign any intervention, this reduction was not statistically significant. 
\end{itemize}

%% file: s_related.tex
\section{Related Work}

There is an increasing recognition in the spaces of machine learning and social computing that accounting for and addressing uncertainty in crowd judgments is an important problem to tackle. In this section we will review this body of prior work, focusing on: (1) establishing the distinction between error and uncertainty; (2) understanding how (aggregate) measures of uncertainty have been utilized in existing systems and workflows; (3) exploring some theoretical frameworks around distinguishing sources of uncertainty; and (4) discussing prior work around context and deliberation and how they informed the design of the interventions we will be using to address the sources of uncertainty.

\subsection{Error v.s. Uncertainty}

Uncertainty observed in group judgments in crowdsourced contexts has historically been attributed to \textit{errors}. As a result of the time-constrained nature of crowdsourced tasks and the limited expertise and attention of crowd workers, errors can certainly often occur~\cite{snow-etal-2008-cheap,Panagiotis-2010-crowd-qc}. Much of the prior work around reducing uncertainty have thus focused on reducing the occurrence and impact of \textit{errors} to the judgment results. One view attributes errors to deficiencies in instructions, leading to confusion and consequent errors on the task~\cite{Wu2017ConfusingTC,Papoutsaki2015CrowdsourcingFS}. Others have pointed to how better training~\cite{gadiraju2015training} or selection~\cite{liu-naacl2016-gating} of workers can reduce error on complex tasks that are difficult to instruct through text and examples alone. Further works have also examined how workers can be kept attentive~\cite{bernstein-2011-realtime} to maintain quality \textit{throughout} a longer task~\cite{Hettiachchi-2021-variable-gold}.
From the labor incentive side, some have viewed the problem of \textit{errors} as a result of adversarial intent (i.e., spam)~\cite{Passonneau2013TheBO,vuurens2011much}, and thus propose incentive-based solutions~\cite{difallah2012mechanical}.
Finally, a data (rather than task) focused view poses the idea of utilizing models to correct for errors post-hoc, inferring worker quality from redundant annotations~\cite{Dawid-1979-MaximumLE} or performance on gold standard questions with known answers~\cite{wiebe-etal-1999-development}. 

However, this \textit{error}-centric view of uncertainty becomes limited when applied to a more recent demand for human judgment---deciding on cases where there may not be a common understanding of what is \textit{ground truth}. Indeed, applying error reduction mechanisms to complex, socially-situated, and even subjective judgments has resulted in problems with downstream models~\cite{Geiger2020-GarbageIG}. 
Problems like demographic skew~\cite{difallah2018demographics} or cognitive biases~\cite{Hilbert2012TowardAS} cannot be ``patched out'' in the same way as errors.
In fact, in many such situations, even the introduction of expertise and removal of time constraint does not fully mitigate all uncertainty~\cite{pavlick-kwiatkowski-2019-inherent,Dumitrache2018-medical-relation-extraction,aroyo2013crowd}.

The above has led to this work, where we recognize that while \textit{errors} are still a part of observed judgment uncertainty, an increasingly important challenge also falls upon how to account for sources of uncertainty not caused by annotator \textit{error} but rather arise from other factors of the task, like ambiguous cases~\cite{Beyer2020AreWD} or legitimate disagreements due to different perspectives and backgrounds of annotators~\cite{sap2019-risk}. Consequently, this work doesn't seek to replace traditional error mitigation, but instead serves to improve how we address \textit{non-error} sources of uncertainty.

\subsection{Accounting for Uncertainty in Human Judgments}

The most straightforward view of uncertainty today, is to treat it as a filter---if the annotators don't agree, then reliable judgment failed. 
Following this view, solutions focus on measuring and evaluating disagreement, such as through inter-rater reliability metrics~\cite{Hallgren2012ComputingIR}.
If agreement is too low, data may be discarded, or more annotators brought onboard, until the agreement becomes satisfactory~\cite{Russakovsky-2015-ImageNetLS,zhang-etal-2017-tacred}.
Achieving a certain threshold of annotator agreement is thus used as a certificate of the quality of a dataset~\cite{snow-etal-2008-cheap}. 
However, as has been observed, the instances to be judged can themselves lie on a spectrum in terms of how certainly they can be judged~\cite{swayamdipta-etal-2020-dataset,welty2019-metrology}, so dropping examples or over-sampling annotators could lead to biases that may then be overlooked due to limited transparency in the dataset construction process~\cite{Gebru-2021-DatasheetsFD}.

Measurement or direct elicitation of uncertainty has also been proposed as a way of building out the task itself.
With unclear guidelines or under-specified tasks being collaboratively refined through processes involving the annotators too~\cite{manam2018wingit,k2019taskmate,bragg-2018-sprout,chang2017revolt}.

One final view on uncertainty focuses on the final application of the human judgment results, proposing that uncertainty be passed along to downstream models as-is. 
For example, systems have been introduced to learn from uncertain labels~\cite{Yang-2019-LearnTB,Yan-2007-RankingWU} and, within the machine learning field, an increasing body of work seeks to harness dis-aggregated labels as the training data for models that achieve higher performance~\cite{Inel2017-HarnessingDI,fornaciari-etal-2021-beyond}.
The application of uncertain labels is not limited to just training, they have also been used in evaluation~\cite{gordon2021-deconvolution} to produce more realistic assessments of performance.

However, there is likely no one-size-fits-all way to work with uncertainty, and even systems trained with uncertainty data can still fail to be socially cognizant~\cite{birhane2021-automating}. 
Not all sources of uncertainty should be treated the same way, even within the same dataset.
Our work thus engages with the recent recognition that to optimally address uncertainty, we should categorize and separately quantify it.

\subsection{Different Sources of Uncertainty}

The quantification of uncertainty has traditionally been done through a statistical lens, by presenting and inferring a probability distribution~\cite{crowder2020introduction} from the judgments collected.
Along this view, recent works have made efforts to quantify uncertainty through capturing \textit{distributions} over the entire set of responses. 
However, answer distributions are costly to produce~\cite{chung2019efficient} and distributions alone don't necessarily offer insight into what may have contributed to the observed result---leading to the need for further analysis and data collection~\cite{welty2019-metrology}.

Thus, more recently, there have been works that aim at \textit{categorizing} different sources of uncertainty separately.
Inspired by earlier works in statistics around measurement
~\cite{Hora1996AleatoryAE,Kiureghian2009AleatoryOE}, one such framework proposes the distinction of: \textit{aleatoric} (or aleatory) uncertainty---where uncertainty arises from the natural unpredictable variance in the property/phenomena measured, and \textit{epistemic} uncertainty---where uncertainty arises from a the limitations of our models, tools, and understanding~\cite{Hullermeier2019AleatoricAE,Walker2003DefiningUA}. 
However, the practical utility of this formulation of uncertainty has sometimes been criticized too, as our evolving understanding of the problem can mean what was once irreducible uncertainty instead was caused by of newly discovered factors~\cite{fox2011distinguishing}.

In this work, we don't directly adopt this prior categorization as-is, instead posing a modified categorization by looking at uncertainty through the lens of \textit{ambiguity} and \textit{disagreement}.
We elected to use this view as it was more directly aligned to the mechanism of how we collected judgments for our experimental tasks~\cite{chen2021-goldilocks} 
More generally, however, we do also note that our particular formulation of \textit{ambiguity} and \textit{disagreement} isn't meant to be a comprehensive categorization of uncertainty, and depending the judgments and end goals of the task, a different framework for the quantification and classification of uncertainty may be more suitable~\cite{Soden2022ModesOU}, 

\subsection{Providing Context to Disambiguate}

One source of uncertainty in human judgments is attributed to the \textit{ambiguous} nature of what is being judged.
Ambiguity in this sense, could be seen as the result of a lack of sufficient \textit{context} surrounding the case to be judged, and thus an intervention to reduce it lies in adding additional clarity.
In toxicity rating tasks, adding context about parent posts has been shown to affect the outcome~\cite{pavlopoulos-etal-2020-toxicity,waseem-2016-racist} while context can also be necessary for investigating online abuse cases~\cite{Menini-2021-AbuseIC}. 
Tasks in natural language processing have also seen context added to improve performance~\cite{huang-etal-2012-improving}.
Beyond the limited scope of crowdsourced annotation, context is also commonly used to reduce ambiguity in classical settings like in the legal realm, where judgment processes often involve expert-conducted procedures specifically meant to establish context to remove ambiguity~\cite{lind1973discovery}.
In education, effective student performance assessment also involves context through free-form responses and intermediate steps~\cite{Marshman2018TheCO} in addition to a final answer.

While context is very important, knowing what context to acquire ahead of time, though, can be difficult.
For example, in content moderation on Wikipedia, moderators may investigate a variety of factors, such as past behaviors, metadata (like IP addresses), and correlated activity before making a moderation decision~\cite{solorio-etal-2014-sockpuppet}.
Other communities may choose to investigate different types of context, such as the Civil War portrait identification community drawing upon historic documents and reasoning around timelines as context around the veracity of an identity association~\cite{mohanty2019-photosleuth}.
Even classical annotation may demand context, but in this case, it may be additional images to disambiguate occluded areas~\cite{li2018richly}.
As a result, we only want to do extra work to collect uncertainty when we know what is most useful.

\subsection{Resolving Disagreement through Rubrics and Deliberation}
\label{sec:relwork-rubrics-deliberation}

Another common source of uncertainty in human judgments can be attributed to underlying \textit{disagreements} between individual adjudicators of a case. 
Many sources can contribute to these disagreements, ranging from inconsistent interpretation of the task criteria to the diverse backgrounds of adjudicators resulting in different perspectives.

For tasks involving crowd annotation, rubrics~\cite{Gadiraju2017-clarity} have been proposed as an effective tool to convey requirements and resolve confusion about aspects of the task. However, rubrics that can cover all the edge cases can be hard to create even by experts, so prior work has utilized crowd participants to help create rubrics~\cite{Manam2019-TaskMateAM,Pradhan2021-InSO,bragg-2018-sprout} by finding areas of high disagreement and asking for suggested guidelines. Rubrics and guidelines can also be implicit, such as in the form of examples~\cite{Lee2022AnnotationCT} or anchors that allow comparison with prior cases~\cite{chen2021-goldilocks}.

Beyond challenges in creating rubrics, rubrics and guidelines also have limitations when applied. 
Even when expert-created guidelines are used, adjustments and refinements may be necessary after judgments are made to address issues with the original guidelines~\cite{Stoica2021ReTACREDAS}.
Additionally, beyond layperson crowds, groups of experts judging instances may also just disagree on what the criteria should be~\cite{Schaekermann-2019-medical-adjudication}.
In certain higher-stakes domains like education~\cite{singh-2017-gradescope}, medical diagnosis~\cite{blangis2021variations}, or legal judgments~\cite{terrebonne1981strictly}, it is often the case where the expertise of the humans results in existing guidelines not being applied exactly, instead often conditionally overridden or even contested and overturned.
In socially embedded domains, like content moderation, guidelines can also fall out of alignment as distributional properties of the data or adjudicators shifts, such as when social norms shift on online platforms~\cite{suzor2013evaluating,gillespie2018custodians} or in broader society.
In these situations, past judgments and the criteria that they used may be contested by future adjudicators.

Finally, unclear tasks are not the sole source of disagreement. 
Even when task goals and guidelines are clear, annotators with different perspective may still disagree about how to judge an item based on different reasoning perspectives~\cite{kairam2016parting}. 
Prior work to automate and scale up deliberation through crowdsourcing has shown that simple reflection-based approaches can be effective at resolving disagreements~\cite{drapeau2016microtalk,Kriplean2011ConsiderItIS}.
More recent work utilizes synchronous deliberation~\cite{Chen2019-CiceroMC,Schaekermann2018-resolvable} to provide contextual deliberation where those participating in deliberation can quickly form targeted arguments for the particular points of disagreement.
Some have also examined the trade-offs between various forms of deliberation design choices (such as the participants, deliberative process, communication medium, etc.) and found that effective deliberation involves building an environment that best matches the task~\cite{davies2013online}. 
Of course, more broadly, the successful use of deliberation to resolve disagreement can also depend on other factors.
For example, it can be important to make sure deliberation participants are trained to argue effectively~\cite{Chen2019-CiceroMC} and communicate in a way that is collaborative and inclusive~\cite{cao2021-team-viability}, especially given the conflicting nature of deliberation. 
Additionally, the dynamics of deliberation (an intellective task) as groups also means that it can be important to make sure that those matched into deliberation teams are compatible~\cite{whiting2019-team-fracture}.

In our work, we draw from existing literature on disagreement resolution to build out one of our interventions---deliberation. However, designing the right deliberation can be challenging and depends on the participants and tasks, so our application of a simpler form of deliberation may very likely be less effective than customizing deliberation for the tasks involved.

%% file: s_design.tex
\section{Design}

In this section we will describe our design of the \sys~workflow. Our workflow consists of the following procedure (as also illustrated in Fig.~\ref{fig:workflow-overview}):
\begin{enumerate}
    \item Collect judgments on each instance from individuals in the group using a process that allows measurement of both individual ambiguity and group disagreement for each instance.
    \item Compute two scores for each instance: Ambiguity ($M_{\text{a}}$) and Disagreement ($M_{\text{d}}$), based on the measurements in the previous step.
    \item For each instance, based on its ambiguity and disagreement measurements, assign potential interventions:
    \begin{itemize}
        \item If ambiguity score is above a set threshold, the instance is assigned the \textsc{context} intervention. Under this intervention, additional context is gathered for the instance and incorporated into it.
        \item If disagreement score is above a set threshold, the instance is assigned the \textsc{deliberation} intervention. Under this intervention, a new group is recruited to re-annotate the instance and then conduct deliberation focusing on their disagreements on the judgment for the instance. At the end of the deliberation for each instance, the group will then collectively produce a suggestion for a new general guideline that they think best resolves the disagreement.
    \end{itemize}
    \item Incorporate the new information produced from the interventions. Additional context acquired is included as part of the corresponding instance. Additional guidelines produced are included into the judgment task definitions.  
    \item Repeat the process as necessary until an acceptable level of uncertainty is reached.
\end{enumerate}

In the remaining parts of this section, we will go into more detail about each aspect of the workflow design. 

\begin{figure}[t]
    \centering
    \centering
    \begin{subfigure}[t]{0.47\linewidth}
        \centering
        \includegraphics[scale=0.25]{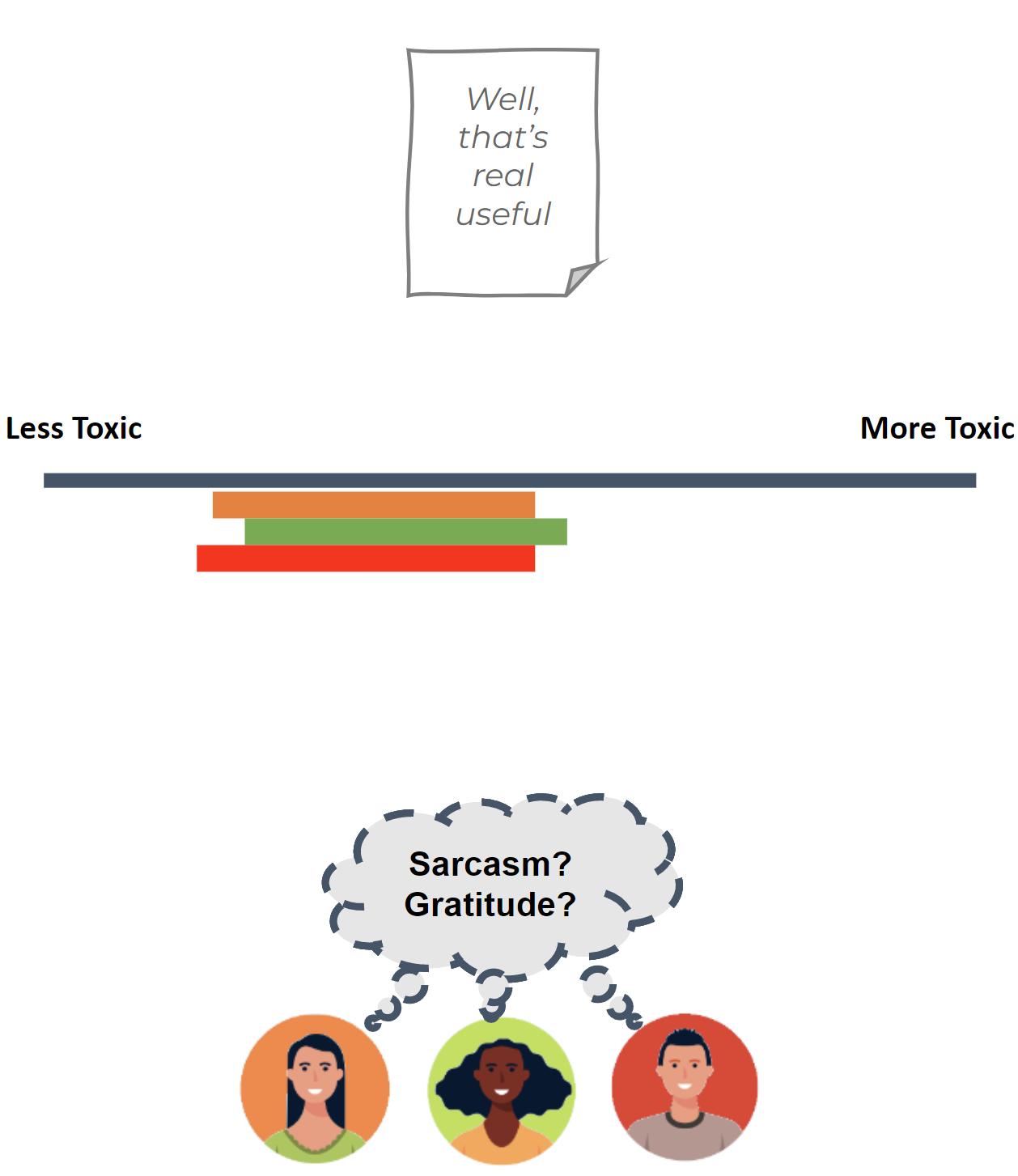}
        \captionsetup{width=.9\linewidth}
        \caption{When annotators find instances \textit{ambiguous}, wider ranges---reflecting more rating levels they find acceptable---are produced.}\label{fig:illust-goldilocks-ambiguity}
    \end{subfigure}
    \begin{subfigure}[t]{0.47\linewidth}
        \centering
        \includegraphics[scale=0.25]{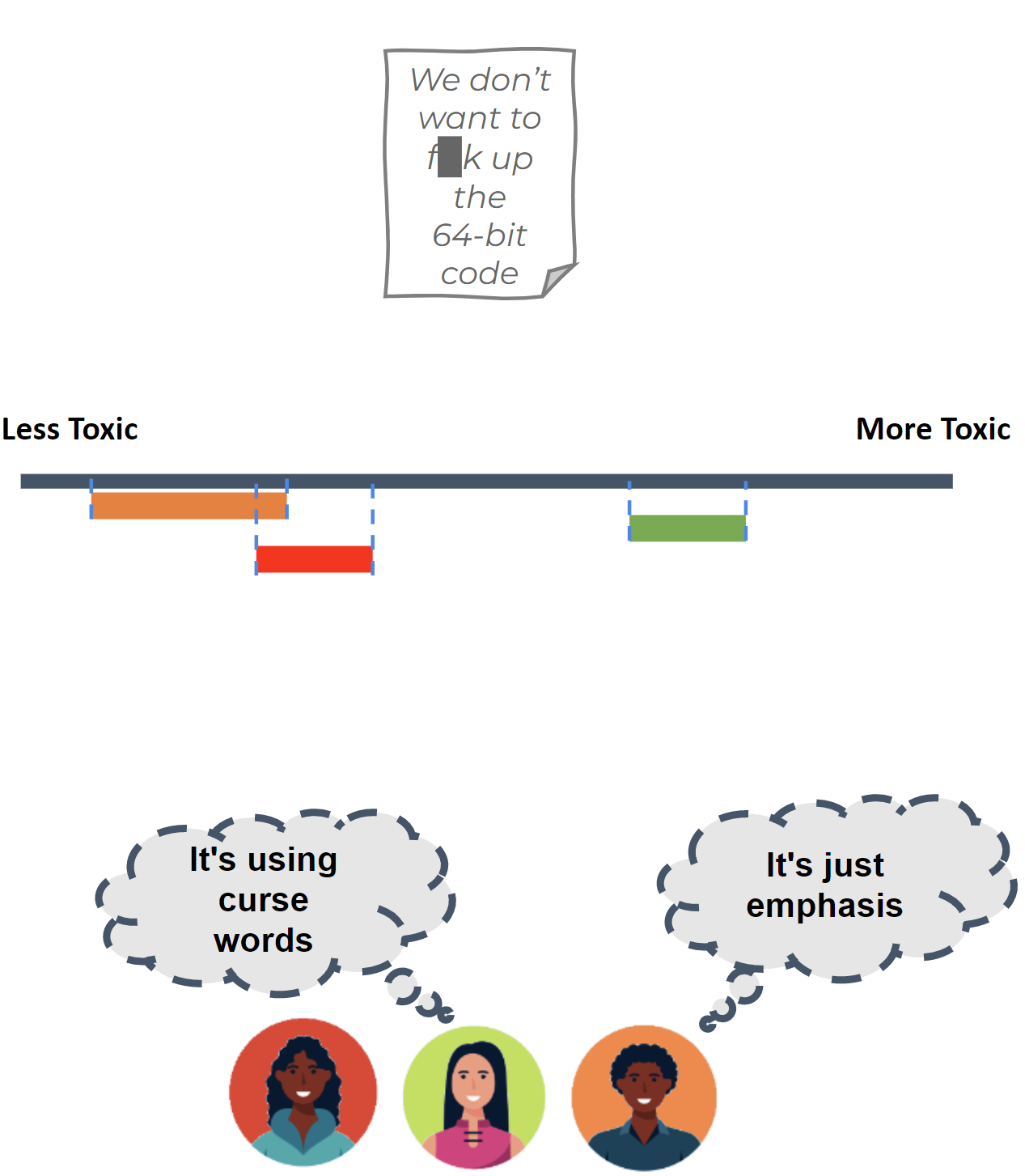}
        \captionsetup{width=.9\linewidth}
        \caption{When annotators \textit{disagree} about the rating, we will see their ranges be placed in different locations on the scale, resulting in less overlap.}\label{fig:illust-goldilocks-disagreement}
    \end{subfigure}
    \caption{Illustration showing the how the two sources of uncertainty---ambiguity and disagreement---can manifest in the form of range measurements produced by a range-based rating annotation tool like Goldilocks~\cite{chen2021-goldilocks}.}
\label{fig:illust-goldilocks}
\end{figure}

\subsection{Measuring Sources of Uncertainty}

\begin{figure}[t]
    \centering
    \includegraphics[scale=0.33]{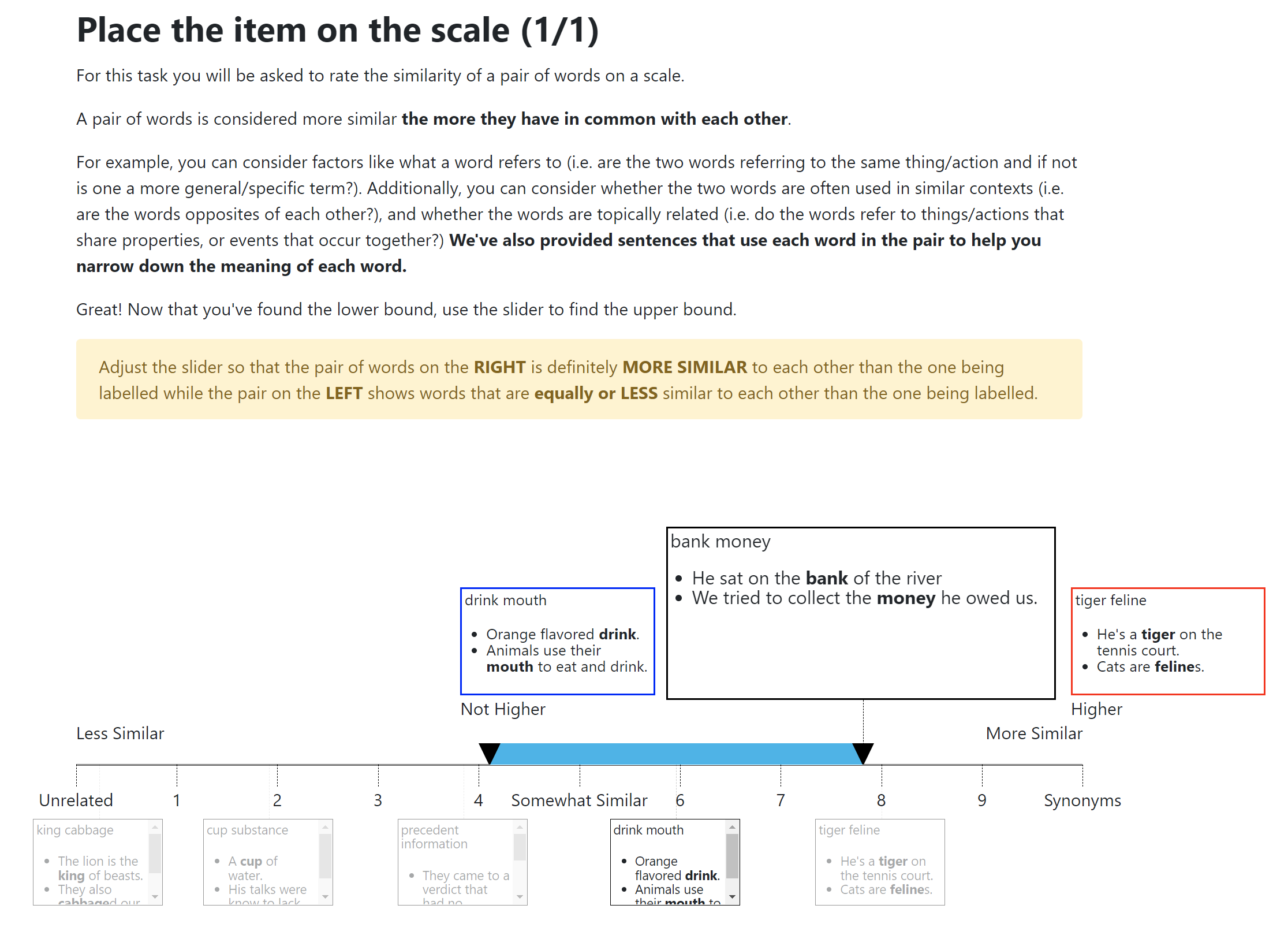}
    \caption{A screen capture of the interface used in the annotation process. This annotation tool allows us to collect measurements of individual judgments by annotators of their observed ambiguity of each item and allows us to measure disagreement through comparing the ranges across different annotators.}
\label{fig:annotation-ui}
\end{figure}

In order to select the right intervention, we first need an approach to understand what sources may be contributing to the group's current uncertainty on each instance. In our workflow, we focus on two distinct sources of uncertainty: the amount of ambiguity inherent to each judgment ($M_{\text{a}}$) and the amount of disagreement between judgments ($M_{\text{d}}$). On a high level, one way to think about the distinction between these two sources of uncertainty is through who contributes to the uncertainty: Ambiguity reflects each individual annotator's certainty about their judgment of the item directly collected through our annotation interface (Fig.~\ref{fig:annotation-ui}); Disagreement is an emergent property that results from aggregating judgments across the individual annotators.  

In our experiments, we look at a common application of our workflow in the context of making rating judgments on a continuous scale. 
Before we can apply targeted interventions, we first need an annotation approach that allows us to distinguish different source of uncertainty.
Most common annotation tools that focus on rating judgments measure uncertainty through aggregation, utilizing disagreements as a proxy for uncertainty~\cite{welty2019-metrology,finkelstein2002-wordsim}.
However, in order to apply effective interventions, we need an approach that is able to distinguish the sources of uncertainty.
Some prior work have incorporated means for individual annotators to indicate their confidence through directly providing estimates of their own uncertainty~\cite{chung2019efficient}, however, humans are generally not good at making these types of assessments~\cite{Tversky1974JudgmentUU}.
For our specific application of scalar rating, we make use of an annotation method introduced by prior work, Goldilocks~\cite{chen2021-goldilocks}, which proposes a way to separately collect measurements on the sources of uncertainty evaluated by each annotator individually during their annotation process. 
Goldilocks achieves this by adapting rating judgments as a range annotation task where instead of single ratings, raters produce a range ($[l_i^{\text{(x)}}, u_i^{\text{(x)}}]$) that reflects values that they find acceptable to place the item. 

Using the range annotations collected through this approach, we can define two metrics that quantify different sources of uncertainty for each instance ($x$). 
We first look at ambiguity---the situation where an individual annotator is unsure about the rating of the instance being judged.
With the range-based annotation procedure, we can see that this kind ambiguity would be reflected through the size of the range produced, with ``wider'' ranges corresponding to more ambiguity around the rating (Figure~\ref{fig:illust-goldilocks-ambiguity}).
Thus we can define an ambiguity score for each instance to be the average size of all ranges collected from the group of annotators participating in the judgment process.
\begin{align*}
    \text{Ambiguity} (x, i) &= u_i^{\text{(x)}} - l_i^{\text{(x)}}\\
    M_\text{a} (x) &= \frac{1}{|N|} \sum_{i \in N} \text{Ambiguity} (x, i) \\
\end{align*}

As for (dis)-agreement between participants, we can see that when range-based annotation is used, the more annotators agree, the more likely it is that the ranges they produce will overlap.
So a natural metric can be formed by looking at the amount---in this case the \textit{ratio}---of an annotators range that overlaps with that of another (Figure~\ref{fig:illust-goldilocks-disagreement}).
However, unlike with range sizes which relate to the fixed scale, simply computing the overlap would result in a metric that is also affected by the size of the ranges (or in our case, the \textit{ambiguity}).
We can see that as the absolute size of any of the ranges increases (reflecting higher \textit{ambiguity}), the likelihood of that range to overlap with another also increases, resulting in a higher overlap ratio.
To account for this and derive a metric for disagreement, we don't directly use the overlap ratio, but instead compare the difference between the measured overlap ratio and the \textit{expected} overlap ratio given the size of the ranges being compared. 
We note that for any range $[l, u]$, the expected overlap ratio of it compared to another uniformly randomly placed range $[l', u']$ is equal to the size of the other range $(u' - l')$.
Given the observations above, we define (dis)-agreement as:

\begin{align*}
    \text{Overlap}(l, u, l', u') &= \max(\min(u, u') - \max(l, l'), 0) / (u - l)\\
    \text{Agreement} (x, i) &= \sum_{j \neq i \in N} \text{Overlap}(l_i, u_i, l_j, u_j) - (u_j - l_j)\\
    M_\text{d} (x) &= - \frac{1}{|N|} \sum_{i \in N} \text{Agreement} (x, i)
\end{align*}

Intuitively, higher agreement scores for an annotator on an instance would indicate more agreement between that annotator and their peers. 
Positive scores imply that the agreement on this instance was higher than random---that annotators leaned towards \textit{agreement}, while negative scores indicate lower than random agreement---that the annotators leaned towards \textit{disagreement}. 
We note that such a definition of agreement for each annotator is generally not commutative (i.e., the agreement between a pair of annotators A, B measured from A is not necessarily equivalent to that measured from B). This reflects the natural asymmetry present in agreement as exposed through range overlap---for a hypothetical pair of annotators, the one with a ``narrower'' (subset) range may agree with their ``wider'' (superset) partner as both accept the ratings in the ``narrow'' range, while from the partner's perspective some ratings that they indicated as acceptable were not accepted by their ``narrower''-ranged partner.
Finally, to make the metric intuitive, we can take the negation of the ``agreement'' metric to define \textit{dis}agreement. We can arrive at a per-instance disagreement score by taking the average disagreement across all annotators. 

\subsection{Gathering Additional Context}

In cases where ambiguity is high among individual judgments, \textit{context} has been shown to be an effective way to reduce this uncertainty in both traditional human judgment settings~\cite{Schaekermann-2019-medical-adjudication} as well as for group judgments facilitated in the form of crowdsourced tasks~\cite{Menini-2021-AbuseIC}. 
However, depending on the judgment task involved, \textit{context} itself can encompass a wide variety of types of information, all of which come with varying amounts of cost involved to capture while not necessarily proving effective for reducing ambiguity.
Even when capturing context is cheap, presenting too much context can risk exhausting the limited attention capacity of human adjudicators and bog down the judgment process~\cite{Rosenhan1994NotetakingCA} and misleading context could result in bad decisions~\cite{Thorley2020MisinformationED}.
As a result, if we want to have human judgments that are scalable, it is likely that attempting to \textit{comprehensively} capture context will be an intractable goal.
Thus we need to build a process for gathering additional context that can be informed by measurements of what cases are actually ambiguous and may benefit from context.

In \sys, we formulate the collection of context as an open-ended process that is customized depending on the particular domain that our workflow is applied on. 
While we can't comprehensively provide guidance for all applications, we will give some brief concrete examples of how context may be collected in a couple of setups: one for a more community-involved task of \textbf{content moderation}, and one for a more annotation-focused task of \textbf{dataset labelling}.

\textbf{Content moderation} in many online communities often takes the form of a committee of \textit{moderators} who collectively decide on a moderation action (such as demoting or removing content, placing a ban on the user, or doing nothing)~\cite{Fan2020-DigitalJA,McGillicuddy2020ControllingBB}. 
While cases may have clear evidence supporting a certain action, historically there have been high-profile cases where limited context contributed to journalistic content being classified as pornographic~\cite{gillespie2018custodians, matthew2016, mullen2016}.
For this type of judgment task, \sys~can identify sets of cases that may require further context due to high ambiguity. The committee can then iterate over such cases to indicate what context might resolve the ambiguity in that case, and deputize a separate group of investigators to collect evidence around the case, who examine logs and metadata and put together a ``casebook''. This investigation ``casebook'' would then become the context attached to a case~\cite{Fan2020-DigitalJA}. In the case of \textbf{dataset labelling}, \sys~would directly inform crowd requesters of ambiguous cases through the ambiguity metric. Crowdsourcing requesters might then launch additional tasks, where workers are asked to augment ambiguous instances with hypothetical context that removes the ambiguity, and reincorporating the clarified cases. Alternatively, if data can be re-collected, such if unclear images~\cite{Deng2009ImageNetAL} resulted from their own data collection, requesters might instead trigger the process to collect data again on that instance, and add the new data as context.

In our experiments, we didn't focus on optimizing any specific context gathering strategy. Instead, we simulate a general context acquisition process by taking a dataset that already has contextual cues collected, and withholding the context to mimic a scenario where the efforts to add the context have not yet been made.

\subsection{Using Deliberation to Resolve Disagreement}

\begin{figure}[t]
    \centering
    \includegraphics[scale=0.34]{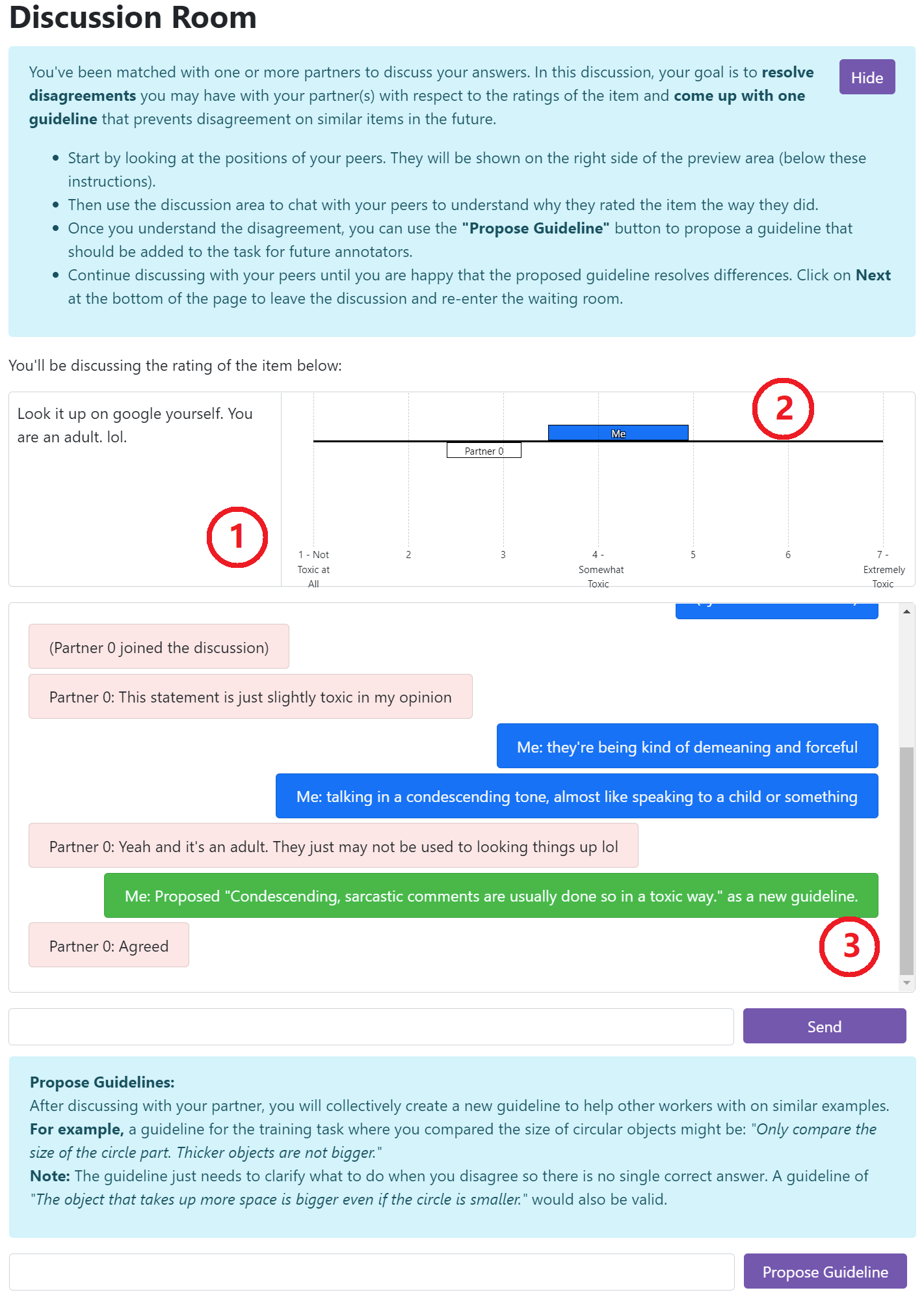}
    \caption{A screen capture of the deliberation interface used in our experiments. There are 3 main components to the interface: (1) A preview of the instance that was rated, (2) A visualization of the range answers of each participant shown on the same scale, and (3) The synchronous discussion area. }
\label{fig:deliberation-ui}
\end{figure}

The design of our deliberation process is inspired by prior work on resolving disagreement using synchronous deliberation~\cite{Schaekermann2018-resolvable,Chen2019-CiceroMC}. In our workflow, the disagreement metric $M_{\text{d}}$ is used to automatically \textit{find} candidate instances that may benefit the most from deliberation---cases where disagreement is the primary source of uncertainty. Then a group deliberates on each example by first independently performing a judgment on the item, and then collectively discussing synchronously. Judgments from each group member is visualized during the deliberation process and the group is collectively prompted to use this to compare their own judgment to those of their peers. Deliberation participants are prompted to consider and elaborate to peers the criteria they used to make their judgment. However, unlike in traditional deliberation systems where the outcome of the deliberation is a judgment on the instance, the goal of our deliberation process is to produce a generalizable guideline for resolving similar disagreements. After engaging in the discussion-based deliberation process, participants are prompted to consider the perspectives they observed during deliberation as well as the deliberation outcome to collaboratively propose a guideline for future examples that \textit{resolve}s the difference in perspective for this instance. 

After all the deliberations have concluded, proposed guidelines can be collected and, if needed, de-duplicated. This produces a final set of guidelines that can be incorporated back into the task, so that future disagreements of a similar type are accounted for in the task itself. We note that this overall workflow is also reminiscent of prior methods proposed to utilize worker-provided feedback to improve the quality of instructions in crowdsourcing tasks~\cite{Manam2019-TaskMateAM,Pradhan2021-InSO}. However, our workflow makes use of the deliberation process to focus the participants on proposing more effective resolutions that account for the disagreements observed rather than inadequate instructions.

\subsection{System Prototype Implementation}

In this section we describe some technical details around the prototype\footnote{Code available: \url{https://github.com/Social-Futures-Lab/targeted-interventions-code}} that we used to conduct our experiments. 
To build out the system prototype for our workflow, we created 2 main components: (1) an \textbf{annotation} application to collect range-based scalar ratings enabling the measurement of ambiguity and disagreement (Figure~\ref{fig:annotation-ui}); and (2) a \textbf{deliberation} application that collects range-based ratings and then matches participants into synchronous deliberation sessions (Figure~\ref{fig:deliberation-ui}).

Our \textbf{annotation} application follows the general design of Goldilocks~\cite{chen2021-goldilocks}, and is implemented as a static web application with the input annotation data and output annotator responses stored directly through Amazon Mechanical Turk (AMT). We use a custom JavaScript toolkit\footnote{Code available: \url{https://github.com/jmchn1994/amt-shim-template}} to interface with AMT and coordinate the experiment conditions and data storage.
Our \textbf{deliberation} application is inspired by the design of prior synchronous deliberation systems~\cite{Schaekermann2018-resolvable,Chen2019-CiceroMC}. Our front-end interfaces with AMT in a way similar to our annotation application, with the deliberation involving the front-end continuously polling for new messages. We coordinate the matching of participants into synchronous discussion rooms with an additional Python-based back-end that also stores the discussions. The matching of discussion participants is guided by a human operator synchronously monitoring an internal facing dashboard (Section~\ref{sec:experiment-deliberation}).

%% file: s_experiments.tex
\section{Experiments}

To evaluate the effects of interventions on group judgment uncertainty, we conducted annotation experiments to collect measurements on the uncertainty of group judgments both before any interventions were conducted and after each intervention was applied. For each instance annotated, we collected ambiguity and disagreement measurements using the our range based annotation application under the following conditions: \textsc{baseline}---no intervention applied, \textsc{context}---context was included as a part of each instance, and \textsc{deliberation}---additional guidelines from the deliberation intervention were provided as part of the task.

\begin{figure}[t]
    \centering
    \begin{subfigure}[t]{0.47\linewidth}
        \centering
        \includegraphics[scale=0.4]{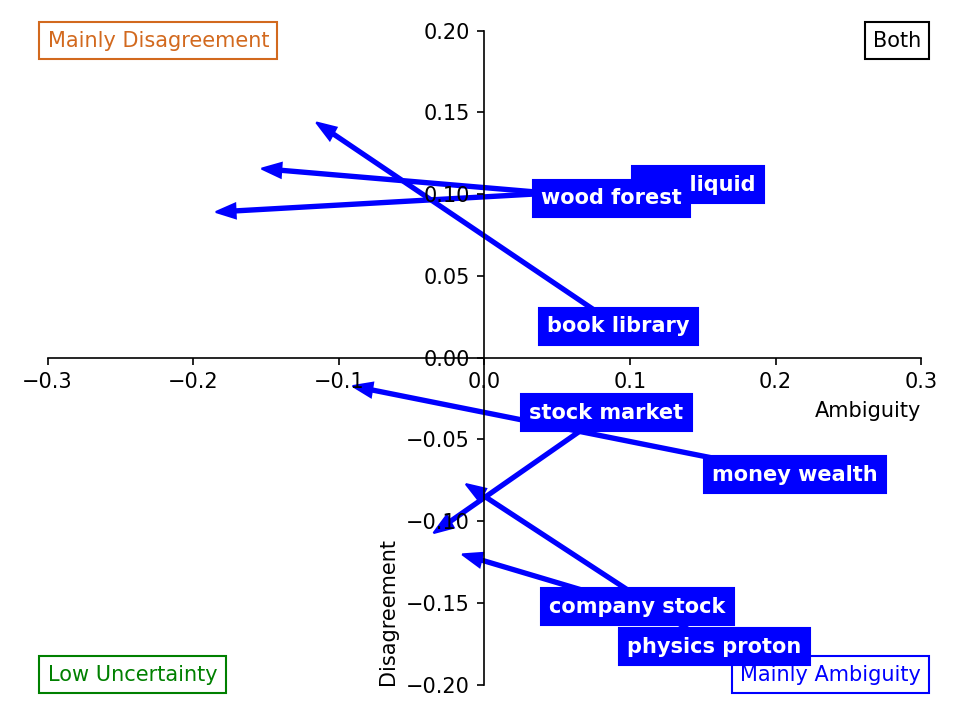}
        \captionsetup{width=.9\linewidth}
        \caption{A sample of items primarily exhibiting ambiguity (blue) and their new uncertainty after applying the \textsc{context} intervention.}\label{fig:illustration_experiment_outcomes_ambi}
    \end{subfigure}
    \begin{subfigure}[t]{0.47\linewidth}
        \centering
        \includegraphics[scale=0.4]{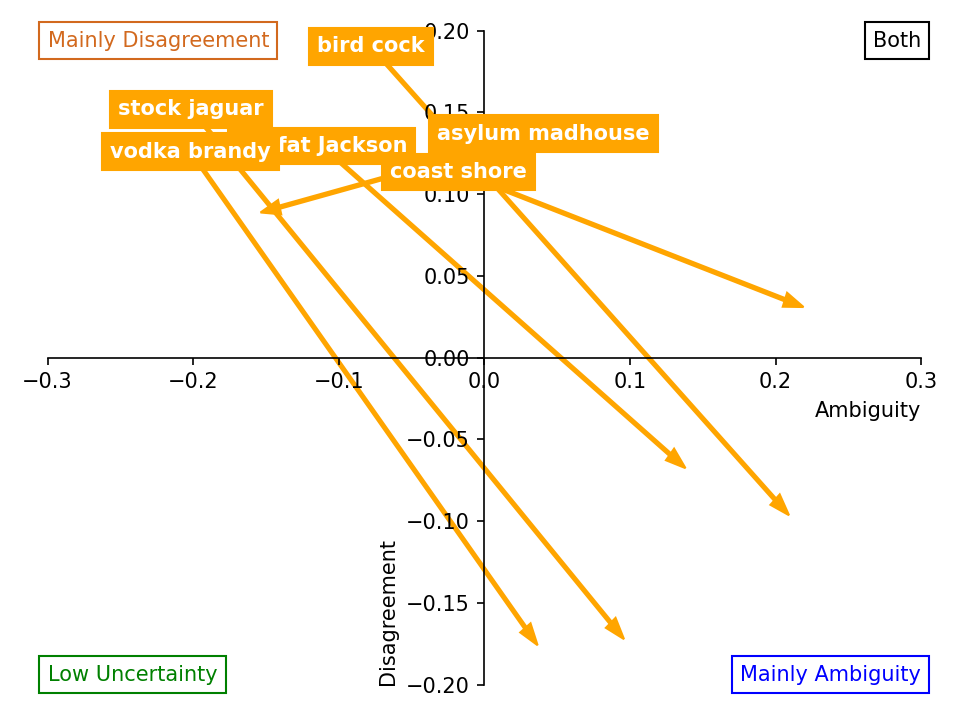}
        \captionsetup{width=.9\linewidth}
        \caption{A sample of items primarily exhibiting disagreement (orange) and their new uncertainty after applying the \textsc{deliberation} intervention.}\label{fig:illustration_experiment_outcomes_disag}
    \end{subfigure}
    \caption{An illustrated figure showing how the uncertainty of a small sample of items moved within the uncertainty space. Items indicated in orange exhibited primarily disagreement. Items indicated in blue exhibited primarily ambiguity. Arrows point to the new location in the uncertainty space after applying the targeted intervention. 
    Scores are re-scaled such that the origin (0, 0) represents the average ambiguity and average disagreement across all items. Positive values indicate above average uncertainty score measurements.}
\label{fig:illustration_experiment_outcomes}
\end{figure}

\subsection{Tasks}

For our experiments, we selected two annotation-based tasks that commonly produce uncertainty in group judgments: word similarity (\textbf{wordsim}) and toxicity rating (\textbf{toxicity}). Both task domains have seen use in prior work and are examples of tasks that contain multiple sources of uncertainty during judgment.

The \textbf{wordsim} domain consists of examples based on an the WordSimilarity-535 Test Collection~\cite{finkelstein2002-wordsim} and is structured as a task to judge the relatedness of pairs of words on a 0-10 scale. This domain was selected because it features varied sources that contribute to uncertainty of both the group and individuals. For one, the ``relatedness'' of words as a concept is only vaguely defined in the \textbf{wordsim} task itself, which can lead to different notions of the relatedness between different people reflected as different schools of thought such as comparing the relatedness of words through various facets such as their meaning, usage, generality and occurrence patterns. Additionally, many of the words involved in this task have multiple word senses. Because no context is provided to disambiguate which word sense is implied, individual annotators must also decide how to reconcile the ambiguity resulting from possible word senses. To seed the range-based annotation process, we used the existing similarity annotations to select 5 seed word pair examples that were evenly spaced along the range with the lowest variance. We then assembled our annotation dataset by selecting a random subset of 50 word pairs divided into 5 groups of 10 from the remaining items. 

The \textbf{toxicity} domain consists of comments collected from a Wikipedia Talk Pages~\cite{pavlopoulos-etal-2020-toxicity} and is structured as a task to judge the toxicity of each individual comment on a continuous rating scale with 7 point semantic differential scale labels. Judging toxicity itself is a task that comes with considerable uncertainty and disagreement. We note that prior work has shown that the background of each annotator and the circumstances in which comments are posted can greatly affect whether the annotator will see the same post as more toxic or not~\cite{sap2021-annotators}. This gives rise to natural disagreement and ambiguity in annotations. As some comments can be many paragraphs long greatly increasing annotation effort, we first filtered the dataset to select only instances where neither the comment or parent comment exceeded a length of 280 characters. Then, to seed the range-based annotation process, we used the existing toxicity annotations from the dataset source to selected 5 seed comment examples that were evenly spaced along the range with the lowest variance. We then created our annotation dataset by selecting a random subset of 50 comments divided into 5 groups of 10 from the remaining items. 

\subsection{Acquiring Context}

The process of acquiring context in general is usually dependent on the specific goals of the group and the task. As this process is separate from the workflow itself and we did not seek to evaluate the quality of context acquired, we instead simulated the process of acquiring context by using task datasets that were already augmented with context. During annotations in the \textsc{baseline}, context of each item was withheld from the annotators, while it was made available during the \textsc{context} condition. 

For the \textbf{wordsim} task, we took inspiration from prior work~\cite{huang-etal-2012-improving}, which used example sentences that contained the word as a way to provide context. For each word in our dataset, we constructed its context by drawing an example sentence that made use of the word in the same form as it appears in the \textbf{wordsim} pair. These example sentences were drawn from WordNet~\cite{miller-1995-wordnet} when available and when examples were not available, an online dictionary service\footnote{\url{https://www.merriam-webster.com/}} was used. When multiple word senses existed, a random one was selected to draw the example sentence from. Shorter example sentences were prioritized with long sentences manually simplified. Context for each word pair was then constructed by appending the example sentence for each word involved in the pair.

In the case of the \textbf{toxicity} task, our dataset source~\cite{pavlopoulos-etal-2020-toxicity} already contains context information provided in the form of the parent comment of each comment. Context was provided to the annotators by appending the parent comment associated with the item along with a label indicating that it was the parent post.

\begin{figure*}[t]
    \centering
    \begin{subfigure}[t]{\textwidth}
        \centering
        \begin{subfigure}[t]{0.49\textwidth}
            \includegraphics[scale=0.5]{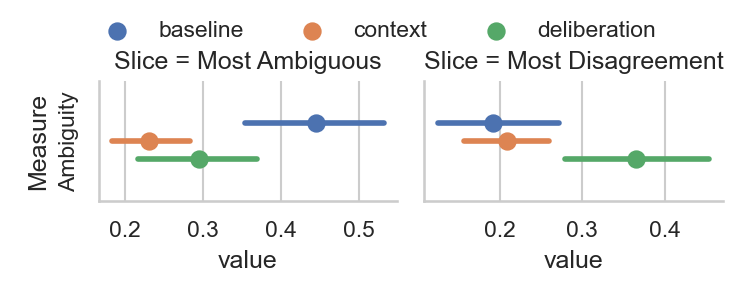}
            \caption{Ambiguity for each slice on the \textit{wordsim} task}\label{fig:intervention_results_ws_ambi}
        \end{subfigure}
        \begin{subfigure}[t]{0.49\textwidth}
            \includegraphics[scale=0.5]{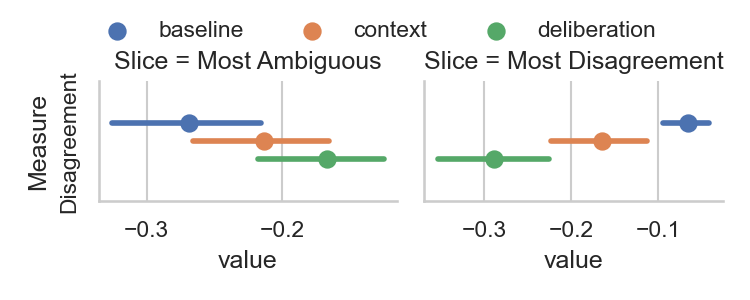}
            \caption{Disagreement for each slice on the \textit{wordsim} task}\label{fig:intervention_results_ws_dis}
        \end{subfigure}
    \end{subfigure}
    \par\bigskip
    \begin{subfigure}[t]{\textwidth}
        \centering
        \begin{subfigure}[t]{0.49\textwidth}
            \includegraphics[scale=0.5]{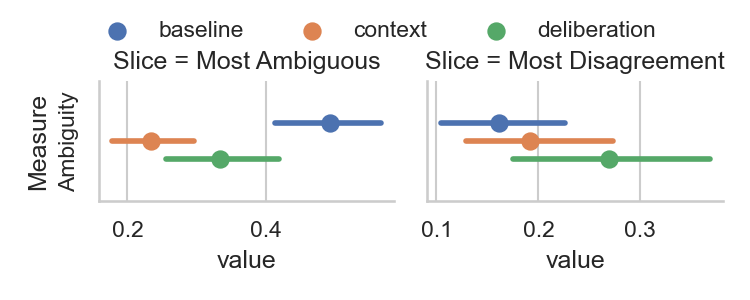}
            \caption{Ambiguity for each slice on the \textit{toxicity} task}\label{fig:intervention_results_tox_ambi}
        \end{subfigure}
        \begin{subfigure}[t]{0.49\textwidth}
            \includegraphics[scale=0.5]{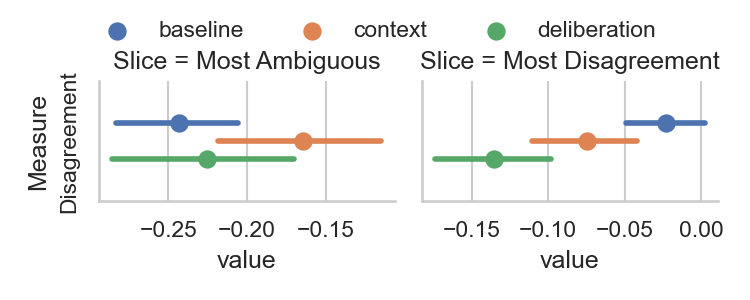}
            \caption{Disagreement for each slice on the \textit{toxicity} task}\label{fig:intervention_results_tox_dis}
        \end{subfigure}
    \end{subfigure}
    \caption{Point plots for each task domain that shows the ambiguity and disagreement measures under the \textsc{baseline}, \textsc{context} and \textsc{deliberation} intervention conditions. For each measure, we look at two slices of the dataset: The instances in the top 10\% by ambiguity $M_a$ (``Most Ambiguous'') and those in the top 10\% by disagreement $M_d$ (``Most Disagreement''). Error bars indicate 95\% confidence intervals.}
\label{fig:intervention_applied}
\end{figure*}

\subsection{Conducting Deliberation}
\label{sec:experiment-deliberation}

We used a crowd task to conduct deliberation to produce guidelines for the \textsc{deliberation} intervention. At the start of the task, each participant first goes through a training session that teaches them to use the annotation interface. After completing this session, participants are placed in a waiting room where they may be assigned either an assessment session or a deliberation session. In an assessment session, the participant uses the range-based annotation interface to provide their judgment for the instance annotated. In a deliberation session, a participant is matched with 1-2 partners and asked to use a real-time synchronous discussion interface (Figure ~\ref{fig:deliberation-ui}) to discuss the disagreement observed in their range annotations and to collaboratively produce a guideline for future annotators. Guidelines can be proposed or updated by any participant and participants may only leave the discussion after a guideline has been proposed. The allocation of assessment and deliberation sessions was done semi-automatically: While a participant is in the waiting room, the deliberation system makes available a set of sessions available to that participant. A deliberation facilitator can then pick among these options to assign to the participant.

Once the deliberation was complete, the final guideline proposals were collected for each item. We then manually de-duplicated proposals by removing those that were similar. Minor modifications were also made to proposals so that they were phrased in a uniform way for each task domain. The proposals collected were then incorporated into the task instructions for the \textsc{deliberation} condition annotation experiments, with 5 new guidelines added to the \textbf{toxicity} task and 6 added to the \textbf{wordsim} task.

\subsection{Recruitment}

We recruited crowd workers from Amazon Mechanical Turk (AMT) to conduct the annotations using an annotation interface based on Goldilocks~\cite{chen2021-goldilocks} for each of the conditions: \textsc{baseline}, \textsc{context} and \textsc{deliberation}. For each condition in each domain, we recruited $25$ workers ($150$ in total). Each participant was given 10 items to annotate for each task deployed. Within each domain, we made sure that a worker could not participate in more than 1 annotation task (displaying a notice and preventing further progression if any tasks beyond the first were attempted), ensuring unique worker pools between conditions in the same task domain. A base payment of \$1.0 was given to participants for completing a training task with another \$1.0 at the end if they completed all annotations. For each annotation completed, participants were paid \$0.3 in the \textbf{wordsim} domain (\$3.0 total) and \$0.5 in the \textbf{toxicity} domain (\$5.0 total). The median hourly pay was measured to be \$13.5 and \$15.9 for the two domains respectively.

Additionally, we also recruited separate AMT workers to participate in deliberation sessions on instances in each domain in order to create the guidelines used in the \textsc{deliberation} condition. For each domain and task group, we recruited 4 discussion participants (a total of 40). We used qualifications to ensure that the workers participating in the deliberation sessions did not participate in the annotations. Workers were paid \$20 for participating in an hour-long discussion task involving 10 discussion and 10 annotation sessions. A bonus of \$4 was given for workers who actively participated in discussions beyond the required 10.

\subsection{Simulation Experiment}

With the annotation experiment data for each of the 3 conditions collected, we are able to simulate the outcome of selecting a targeted intervention for each instance. For our simulation experiment, we used the ambiguity $M_{\text{a}}$ and disagreement $M_{\text{d}}$ scores collected during the \textsc{baseline} condition to decide the intervention to use for that instance.

For our experiments, we selected a threshold value of $0.1$, which targets the instances that ranked in the top 10\% in terms of either ambiguity score or disagreement score. To conduct the simulation, instances were sorted by their $M_{\text{a}}$ and $M_{\text{d}}$ scores collected from the \textsc{baseline} condition. We use this to determine a cutoff threshold for the ambiguity and disagreement scores ($\bar{M_{\text{a}}}, \bar{M_{\text{d}}}$). Then, for each instance in the dataset, we first check its ambiguity score. If $M_{\text{a}} (x) \geq \bar{M_{\text{a}}}$, we assign the context intervention by drawing annotation values from the \textsc{context} condition for this instance and moving on to the next instance. Otherwise, we check the disagreement score, and if $M_{\text{d}} (x) \geq \bar{M_{\text{d}}}$, we will draw annotation values from the \textsc{deliberation} condition for this instance. If neither uncertainty metric was above the threshold, we leave annotation values from the \textsc{baseline} condition unchanged.

\subsection{Results}

To evaluate our workflow, we focused on 2 main aspects: evaluating the effect of each intervention on the type of uncertainty it targets, and evaluating whether dynamically selecting a targeted intervention based on uncertainty measurements for each example can more efficiently reduce uncertainty compared to a uniform application of intervention.

Specifically, we evaluate the following hypotheses:
\begin{itemize}
    \item \textbf{H1-a (Interventions are Effective)}: An intervention is effective at reducing the source of uncertainty it targets: \textsc{context} will be most effective at reducing ambiguity, while \textsc{deliberation} will be most effective at reducing disagreement.
    \item \textbf{H1-b (Interventions are Targeted)}: An intervention is not effective at reducing the type of uncertainty it does not target.
    \item \textbf{H2 (Efficient Uncertainty Reduction)}: A decision process based only on uncertainty measurements collected without any intervention can select a more optimal intervention for each instance that reduces uncertainty more efficiently than a uniform application of an intervention over all instances.
\end{itemize}

\subsubsection{Effectiveness of Targeted Interventions}
\label{sec:results-effectiveness}

In this section, we will examine whether our hypotheses for the effectiveness and targeted nature of interventions is supported in our two task domains. To test our hypotheses, we extract 2 subsets of instances (slices) from each task based on the primary source of uncertainty measured during the \textsc{baseline} annotation. For each domain, we selected the top 10\% instances that had the highest measured ambiguity as a ``Most Ambiguous'' slice and the top 10\% instances that had the highest measured disagreement as a ``Most Disagreement'' slice. Then for each set of instances, we tracked their uncertainty after re-annotation following each intervention (\textsc{context} and \textsc{deliberation}). We visualize these measurements in Figure~\ref{fig:intervention_applied}.

Looking at the slice of ``Most Ambiguous'' instances in each domain, we found that only the \textsc{context} intervention condition was observed to be statistically significant in reducing the ambiguity across both the \textbf{wordsim} and \textbf{toxicity} task domains ($p < 0.001$, observed only between the \textsc{baseline} and \textsc{context} conditions using Tukey's HSD). We found similar results for the slice of ``Most Disagreement'' instances when it came to disagreement, observing only statistically significant reduction in disagreement between \textsc{deliberation} and \textsc{baseline} pairings ($p < 0.001$). This supports \textbf{H1-a} indicating that interventions are effective in reducing the type of uncertainty it targets.

We also examined how interventions affected the other (non-targeted) source of uncertainty. In both the \textbf{wordsim} and \textbf{toxicity} domains we did not observe statistically significant interactions of between the non-targeted condition and \textsc{baseline}. While lack of observing significance does not indicate that the non-targeted conditions had no effect on the source of uncertainty, it does indicate that they are not as effective as the targeted intervention, thus this provides some partial support for \textbf{H1-b}. Curiously, we did find that on the ``Most Disagreement'' slice in the \textbf{wordsim} domain, while \textsc{deliberation} was significant in reducing that disagreement, it also had a significant effect on \textit{increasing} ambiguity. Due to the nature of the task, we hypothesize that the guidelines produced from \textsc{deliberation} resulted in participants considering more factors (word senses, indirect relationships) when determining the relatedness of words and as a consequence of the lack of any other context, they found the instances to be more ambiguous. 

\subsubsection{Efficiency of Decision Process}
\label{sec:results-efficiency}

\begin{figure}[t]
    \centering
    \begin{subfigure}[t]{\linewidth}
        \centering
        \includegraphics[scale=0.6]{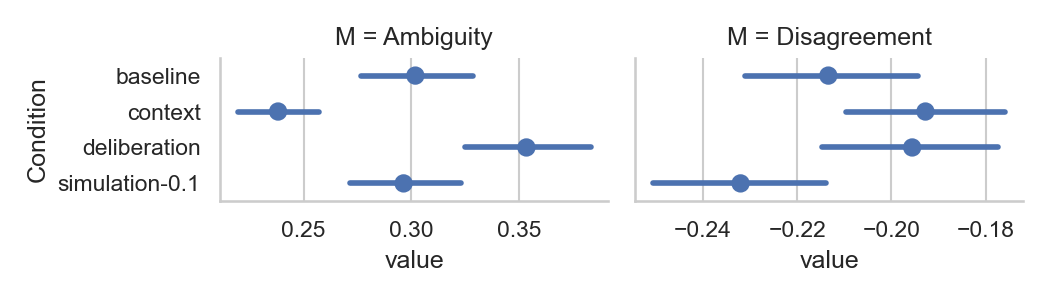}
        \caption{Comparison for the \textit{wordsim} task domain}\label{fig:simulation_results_ws}
    \end{subfigure}
    \par\bigskip
    \begin{subfigure}[t]{\linewidth}
        \centering
        \includegraphics[scale=0.6]{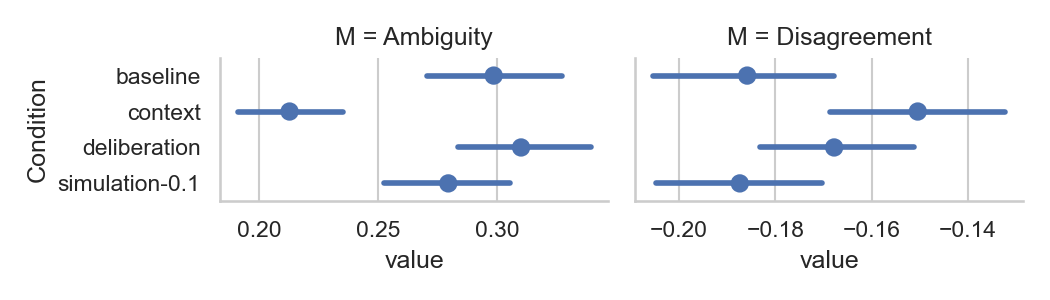}
        \caption{Comparison for the \textit{toxicity} task domain.}\label{fig:simulation_results_tox}
    \end{subfigure}
    \caption{Point plots for each domain that show the ambiguity and disagreement measured after applying a uniform intervention (\textsc{context} or \textsc{deliberation}) across all instances and from simulating the selection of different interventions targeted to each instance \textsc{simulation-0.1}. Error bars indicate 95\% confidence intervals.}
\label{fig:simulation_results}
\end{figure}

Popping up a level and looking at the case of uniformly applying each intervention across the all instances in the entire dataset (Figure~\ref{fig:simulation_results}), we found that \textsc{context} was able to reduce ambiguity in both domains ($p = 0.0027 < 0.01$ and $p < 0.001$ for the \textbf{wordsim} and \textbf{toxicity} domains respectively). However, this seems to also come at a slight cost, also raising the mean disagreement in both cases ($p = 0.026 > 0.01$, not signif.\footnote{We set an a priori significance level at} $p < 0.01$ throughout our statistical tests., for \textbf{toxicity}, $p > 0.01$, not signif., for \textbf{wordsim}). This indicates that applying the same intervention across-the-board to all instances can come with trade-offs, potentially causing increases in sources of uncertainty it was not meant to address. When looking at the \textsc{deliberation} condition, we found no statistically significant effects on either uncertainty source when applied across the entire dataset, with slight increases in the mean value on both measurements. This suggests that while deliberation can be useful for instances with the most disagreement, applying it broadly may be harmful. This result is broadly in line with prior work on deliberation that suggests deliberation is likely only effective when items are already low in ambiguity~\cite{Schaekermann2018-resolvable} and should be used primarily on the challenging high disagreement cases.

Next, we compare our results from the simulated decision process where instances are assigned different interventions based on whether their uncertainty is primarily caused by ambiguity or disagreement. When comparing against the \textsc{baseline}, we found that our simulated process (\textsc{simulation-0.1}) resulted in lower mean values from both ambiguity and disagreement measures in both domains. However, this decrease was not measured to be statistically significant. The lack of significant results is not unexpected, though, as our simulated selection approach only applies an intervention to the top 10\% of instances with highest ambiguity and disagreement as measured during the \textsc{baseline} annotations (only affecting at most 20\% of instances) while all the remaining instances retained their original annotations. We also note that increasing the decision threshold biases results toward the \textsc{context} condition---more significant decreases in ambiguity at the cost of higher disagreement. Interestingly, we observed that our two task domains responded differently to our simulated decision process, with \textbf{wordsim} achieving the most reduction of uncertainty through reducing disagreement (-8.7\%), while \textbf{toxicity} achieved more reduction of ambiguity (-6.4\%). We hypothesize that this may be due to disagreements being more challenging to resolve in \textbf{toxicity} judgments. In the end, while we don't show \textbf{H2} to be true in a statistically significant way with one round of targeted intervention, we do see a differences that may allow us to avoid trade-offs of balancing uniformly adding context or deliberating on all instances.  

%% file: s_discussion.tex
\section{Discussion}

In this section we will first examine the effect of varying the thresholds for selecting interventions and discuss how thresholds (which affect uncertainty reduction on a per-round basis) work in conjunction with iterative improvement style application of our workflow.
Then we will discuss some qualitative observations on the guidelines produced through deliberation and how it may relate to the differences we observe across our two task domains.
Following that, we will discuss how our workflow coordinates situations that involve both ambiguity and disagreement and discuss how our workflow can generalize across different tasks and modalities beyond the crowdsourced scalar rating annotation we used in our experiment.
Finally, we will discuss some of the limitations of the two interventions we explored---context and deliberation---as well as avenues for future work that may resolve some of these limitations.

\begin{figure*}[t]
    \centering
    \begin{subfigure}[t]{\textwidth}
        \centering
        \begin{subfigure}[t]{0.49\textwidth}
            \includegraphics[scale=0.45]{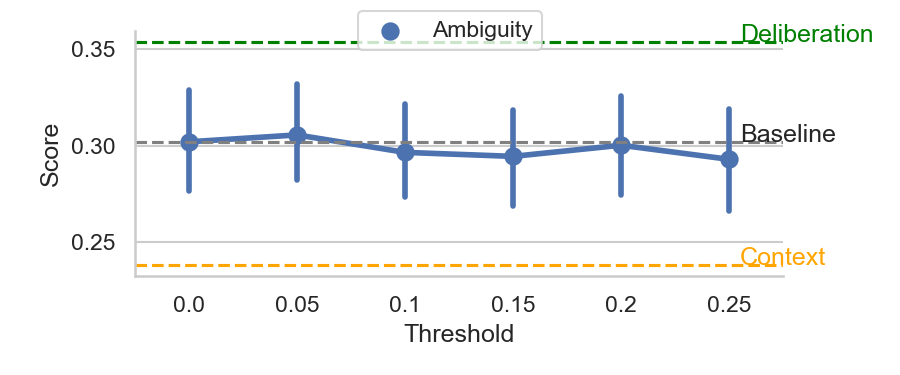}
            \caption{Overall ambiguity on the \textit{wordsim} task}\label{fig:intervention-thresholds-ws-ambi}
        \end{subfigure}
        \begin{subfigure}[t]{0.49\textwidth}
            \includegraphics[scale=0.45]{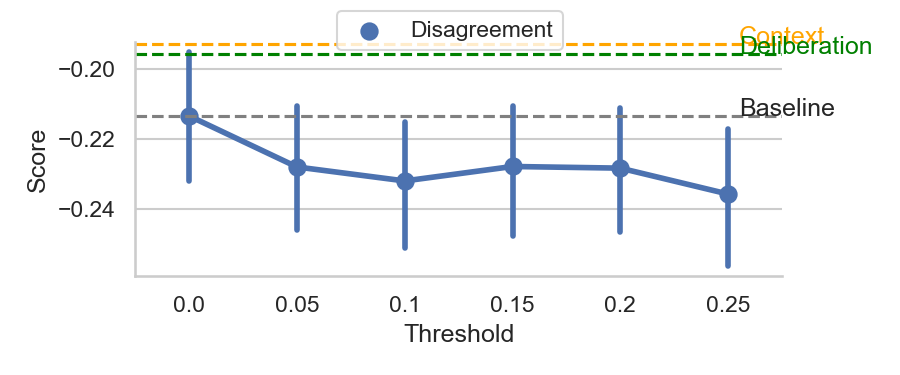}
            \caption{Overall disagreement on the \textit{wordsim} task}\label{fig:intervention-thresholds-ws-disagree}
        \end{subfigure}
    \end{subfigure}
    \par\bigskip
    \begin{subfigure}[t]{\textwidth}
        \centering
        \begin{subfigure}[t]{0.49\textwidth}
            \includegraphics[scale=0.45]{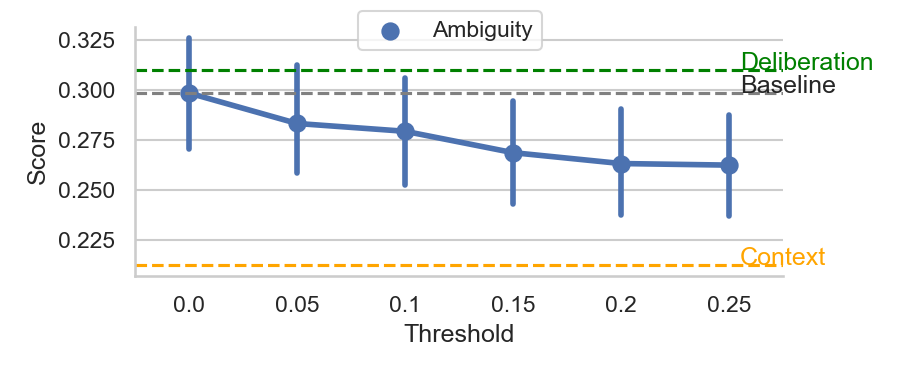}
            \caption{Overall ambiguity on the \textit{toxicity}  task}\label{fig:intervention-thresholds-tox-ambi}
        \end{subfigure}
        \begin{subfigure}[t]{0.49\textwidth}
            \includegraphics[scale=0.45]{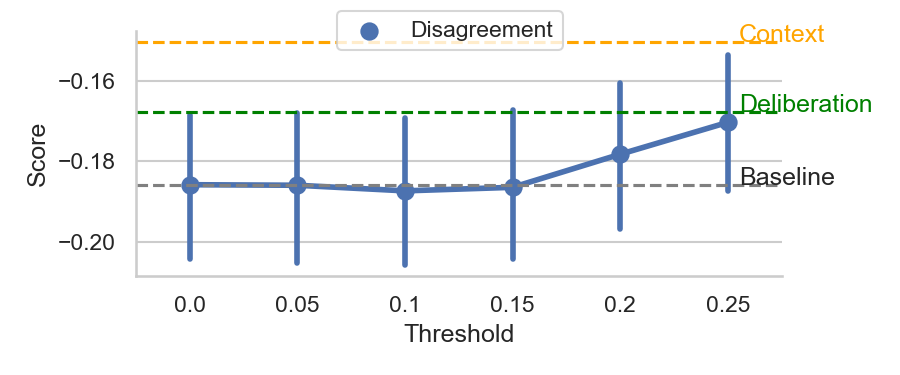}
            \caption{Overall disagreement on the \textit{toxicity}  task}\label{fig:intervention-thresholds-tox-disagree}
        \end{subfigure}
    \end{subfigure}
    \caption{Plots showing the simulated interventions applied at different thresholds of 0\% (no interventions applied), 5\%, 10\%, 15\%, 20\%, and 25\%. For all plots, lower values reflect less uncertainty from the corresponding source. Three reference lines are provided on each graph to indicate the average uncertainty measurements of: \textsc{baseline} (grey), \textsc{context} (orange), and \textsc{deliberation} (green). Error bars indicate 95\% confidence intervals around simulations, confidence intervals for the reference lines are not shown (see Figure~\ref{fig:simulation_results} instead).}
\label{fig:intervention-threshold}
\end{figure*}

\subsection{Intervention Selection Thresholds and Iterative Improvement}

In section \ref{sec:results-efficiency}, we found that we are able to observe reductions in both types of uncertainty by simulating a decision process that applied interventions to the top 10\% of instances with highest ambiguity and disagreement, respectively, though not at a statistically significant level. As at most 20\% of the instances would affected, one question that arises is what happens if we change this threshold to allow interventions to be applied to more (or fewer) instances. 
To explore this question, we adjusted the simulation parameters to simulate the decision process under additional thresholds as shown in Figure~\ref{fig:intervention-threshold}).

Through these simulations, we can observe that the two task domains tested respond differently in terms of their sensitivity to the targeted intervention selected. For the \textit{wordsim} domain, we find that applying targeted interventions reduces \textit{overall} disagreement but achieves relatively little benefit to \textit{overall} ambiguity. 
From our results in Section~\ref{sec:results-effectiveness}, we know that the \textsc{context} intervention is effective at reducing ambiguity for those most ambiguous instances, which indicates that the \textsc{deliberation} intervention likely caused increases in ambiguity on the high-disagreement cases that canceled out the reduction of ambiguity provided by \textsc{context}.
We hypothesize that in this domain, the additional guidelines led to more comprehensive views on ``word similarity'' with annotators realizing that cases they would have been certain about (and thus disagreed with each other on) were actually ambiguous (and that they wouldn't have considered those alternative interpretations had it not been for the guidelines).
On the other hand, for the \textit{toxicity} domain, we find almost the opposite scenario where targeted interventions resulted decreased \textit{overall} ambiguity but had minimal change to (or even increases to) \textit{overall} disagreement.
This suggests that for this domain, more context may have reduced the ambiguity around the setting of the online comments, but may have surfaced new disagreements on what toxicity means for the different annotators~\cite{sap2021-annotators}.

While this simulation result itself is interesting, we note that in practice, one would not be able to find an ``optimal'' threshold using this approach as each intervention would need to be applied to all instances, resulting in an inefficient process. 
Instead, we posit that optimizing the threshold would likely not be the most effective way to reduce uncertainty in practice; rather, a better approach lies in an \textbf{iterative improvement}~\cite{Goto2016UnderstandingCW} formulation where our workflow is run in additional iterations that operate on the data and task after application of the interventions from a previous round.
Prior work has shown that some uncertainty interventions, like deliberation, used in our workflow may only be effective on instances that have low ambiguity and may be counterproductive otherwise~\cite{Schaekermann2018-resolvable,Chen2019-CiceroMC}.
Indeed, we observe this in Figure~\ref{fig:simulation_results}, where we found that uniformly applying deliberation across all instances can slightly increase overall disagreement in both domains.
However, targeted application of deliberation can reduce disagreement even if indiscriminate application does not (Figure~\ref{fig:intervention-thresholds-tox-disagree}).
This suggests that a more effective approach lies in iterating on the workflow rather than optimizing thresholds: after each iteration, instances that were ambiguous (and thus not suitable for deliberation) may now be less ambiguous, potentially opening them up to deliberation as an effective intervention in the next round.
In an iterative construction of the workflow, selection thresholds can instead be seen as a way to control the rate of uncertainty reduction per-round (almost akin to a ``learning rate''). Lower values are more conservative, affecting overall uncertainty less but more likely to avoid interventions cancelling out each others' benefits, whereas higher values reflect a more optimistic view on interventions, increasing the likelihood of failing to reduce uncertainty in a round, but having a larger impact at each step when it works. 

\subsection{Utility of Guidelines Produced}

In our results, we saw that applying deliberation across the entire dataset can result in increases in disagreement even though we also observe that it reduces disagreement for those cases with the highest disagreement. 
To explore this, we qualitatively examined several of the guidelines produced through the deliberation process to examine how they may not have been effective at scaling to more instances.

For the \textit{wordsim} domain, we found that deliberation resulted in guidelines that outlined additional criteria for what would be considered as ``similar'', such as: ``Antonyms (light/dark, good/evil) are similar.'', ``Causal [sic] and effect between words make them more similar.'', and ``Words part of a natural progression are more similar.''. However, while these guidelines would have likely provided more consistent criteria around the word pairs that were deliberated on to produce them, they still leave opportunities for disagreements around applying them---e.g., would a certain word pair be considered a cause-effect pairing or natural progression?
For the \textit{toxicity} domain, we found that deliberation resulted in new guidelines such as the following: ``Statements about policies not people are not considered toxic.'', ``Demeaning or condescending statements are likely to be toxic.''. Like in \textit{wordsim}, these guidelines are also overall rather narrow (``statements about policies'') or could be vague when context was limited (``condescending statements'').

While this is not a comprehensive exploration of the effectiveness of producing guidelines, we note that it provides insight into why guidelines produced by our particular deliberation formulation may not have generalized well in some cases.
We also note that, even though the setup for \sys~in our evaluation uses a specific deliberation design, our implementation of it is meant as a proof-of-concept, and doesn't reflect the most effective setup for deliberation optimized for our task domains. In practice, groups utilizing deliberation may elect to use alternative designs that improve matching quality or involve expert-led processes.

\subsection{Ambiguity and Disagreement All at Once}

As we have observed in our experiments, while ambiguity and disagreement are largely distinct types of uncertainty, it is also not uncommon for an instance to have both high ambiguity and high disagreement. What should one choose to focus on when this occurs? In our simulated version of targeted intervention, we opted to prioritize resolving ambiguity before disagreement. This decision was informed by prior work indicating that the deliberation intervention we used (in the form of self-contained synchronous online discussions) may fail to resolve disagreement in cases with high ambiguity~\cite{Schaekermann2018-resolvable}. 
However, this particular choice may not always be optimal. It may be more productive in some cases to prioritize discussion instead. For example, moderation decisions around developing topics may involve both creating consensus on guidelines (as has been seen in platforms' adaptations to misinformation campaigns related to COVID-19), as well as collecting evidence (current scientific consensus, evaluating whether content is connected to larger misinformation campaigns, etc.) that backs a final decision. 
In these situations, applying deliberation first or in parallel can shed light on the context that will become necessary, ultimately directing a more effective context collection process. 
A promising avenue of future work may be to develop approaches to hybridize the collection of context and the deliberation process, allowing groups to switch back-and-forth between the two as needs arise.

\subsection{Generalizing our Approach Across Different Tasks and Modalities}
\label{sec:discussion-generalizing}

In this work, we evaluated \sys~on the specific judgment modality of continuous scalar ratings for short text-based tasks under a continuous rating scale. 
However, more broadly speaking, there are many more scenarios (e.g., expert involved group judgments) and judgment modalities (e.g., single or multi-label categorical classification) involving group human judgments where it can be beneficial to reduce uncertainty in a targeted way. 
In this section, we will discuss two areas where we anticipate opportunities for generalizing our workflow: supporting \textbf{complex tasks} through a task specific iterative process, and supporting \textbf{modalities beyond scalar rating}.

While \sys~was built around reducing uncertainty in group judgments in a crowdsourced setting, the higher level concepts of classifying and quantifying uncertainty and applying targeted interventions could apply to other types of tasks and scenarios. 
For example, in education settings, a \sys-inspired workflow might involve expert-level teaching assistants using a similar tool during grading to measure disagreements around score assignment or ambiguity surrounding some types of answers.
These measurements might then feed into group discussions that result in rubric refinements (to reduce uncertainty) or future updates to assignment questions (to reduce ambiguity).
Likewise, online communities may include a \sys-inspired process as a part of their moderation process, allowing uncertainty around decisions to audited and addressed as needed.
An iterative version of \sys~for community use might also provide longer term stability as community norms evolve over time~\cite{zhang2017characterizing}.

As for other modalities, like categorical or multi-label annotations, \sys~can be adjusted to use uncertainty categorizations and metrics that are built for those modalities. For example, soft labels~\cite{Collins2022ElicitingAL} could be used in categorical settings, allowing metrics like \textit{dispersion} (e.g., variance) of single-annotator distributions to measure ambiguity and \textit{divergence} (e.g., KL divergence, Wasserstein distance) to measure inter-annotator disagreement.
We also do note that, while the potential to generalize is large, \sys~is still expected to be most effective for reducing uncertainty in more complex or subjective judgment settings, with more perception focused tasks expected to benefit less from a componentized view of uncertainty. 

\subsection{Caveats of Context}

While in general context can reduce ambiguity, in our experiments, we did observe cases where context contributed to an \textit{increase} in uncertainty. These occurred mainly when the context was unexpected. For example, examining the cases where ambiguity increased after context was added in the \textbf{wordsim} domain, we saw instances like ``bank, money'' becoming more ambiguous. However, this did not reflect a failure of the context intervention, rather, we observed that the context introduced (a usage example of ``He sat on the \textbf{bank} of the river'') was likely not a word sense expected by the annotators in the original judgment. Despite increasing uncertainty, this outcome would likely be desirable in practice as it reflected a real increase in uncertainty. 

At a higher level, there do exist limits to how far additional context can go to reducing uncertainty in practice and eventually we are likely to run into diminishing returns of gathering more context for minimal benefit to disambiguation. Thus it may be advisable to keep track of the magnitude of uncertainty reduction should further iterations be invoked, and balance the cost accordingly. 

\subsection{Limits to Scaling Task Specification with Deliberation}

Finally, we also note that there are limits to scaling the current design of our deliberation process. 
In the current process, deliberation produces additional guidelines which are incorporated into the instructions. 
While processes like de-duplication and reorganization can be done by task requesters, as the task specification becomes increasingly precise, the instructions grounding the task itself can eventually become unwieldy~\cite{Wu2017ConfusingTC}. 
We see this in current complex tasks like content moderation, where extensive training is given to paid contractors to apply the exact moderation guidelines.
If guidelines become too complex, their ability to resolve disagreement would be reduced, as people struggle to understand or find a relevant guideline. 

A potential solution to address overly complex task specifications may lie in the realm of legal case building, where decision boundaries can be impossibly complex, and judgments instead often rely on references to past decisions rather than invoking a statutory law or guideline. This opens up possibility of future work to explore alternatives to reducing disagreement beyond deliberation.

%% file: s_conclusion.tex
\section{Conclusion}
In this paper, we present a new workflow for more efficiently reducing uncertainty in group judgments by applying a targeted intervention on each instance based measurements relating to ambiguity and disagreement. Through our experiments, we find that the interventions of adding context and conducting deliberation do most effectively reduce the type of uncertainty it targets. We also observe that dynamic selection of interventions on a per-item bases has the potential to avoid the trade-offs in uniformly applying interventions to all items.